# A Knowledge Base for Arts and Inclusion - The Dataverse data archival platform as a knowledge base management system enabling multimodal accessibility

Moa Johansson[1][0009-0007-4965-9243] , Vyacheslav Tykhonov[2][0000-0001-9447-9830] , Sophia Alexandersson[1][0009-0000-8824-4357] Kim Ferguson[2][0000-0001-6483-1936] James Hanlon[3][0009-0004-2550-2450] Andrea Scharnhorst[2][0000-0001-8879-8798] Nigel Osborne[1][0000-0002-5517-6738]

[1] ShareMusic & Performing Arts, 563 32 Gränna, Sweden
[2] Data Archiving and Networked Services, Royal Netherlands Academy of Arts and Sciences (DANS-KNAW), 2593 HW The Hague, The Netherlands
[3] X-System Ltd, Hampshire, PO157FX England, UK

moa@sharemusic.se

**Abstract.** Creating an inclusive art environment requires engaging multiple senses for a fully immersive experience. Culture is inherently synesthetic, enriched by all senses within a shared time and space. In an optimal synesthetic setting, people of all abilities can connect meaningfully; when one sense is compromised, other channels can be enhanced to compensate. This is the power of multimodality.

Digital technology is increasingly able to capture aspects of multimodality. To document multimodality aspects of cultural practices and products for the long-term remains a challenge. Many artistic products from the performing arts tend to be multimodal, and are often immersive, so only a multimodal repository can offer a platform for this work. To our knowledge there is no single, comprehensive repository with a knowledge base to serve arts and disability. By knowledge base, we mean classifications, taxonomies, or ontologies (in short, knowledge organisation systems)

This paper presents innovative ways to develop a knowledge base which capture multimodal features of archived representations of cultural assets, but also indicate various forms how to interact with them including machine-readable description. We will demonstrate how back-end and front-end applications, in a combined effort, can support accessible archiving and data management for complex digital objects born out of artistic practices and make them available for wider audiences.

# 1        Introduction

## 1.1        New Multimodal Needs for Accessibility and Participation.

Creating an inclusive art environment requires engaging multiple senses for a fully immersive experience. Culture is inherently synesthetic, enriched by all senses within a shared time and space. In an optimal synesthetic setting, people of all abilities can connect meaningfully; when one sense is compromised, other channels can be enhanced to compensate. This is the power of multimodality.

The core of the paper is to explore notions of multimodality in connection with accessibility and long-term archiving. The central research question is "How to respond to new multimodal needs for accessibility and participation?"  More concretely we investigate what kind of knowledge base is needed for such a response. The concrete case is the design of a repository which will document and make accessible a collection of artifacts around performing arts. The spread of the collection ranges from scientific documentation (articles) about research enabling performing arts to actually recording traces of performing arts instances. It is hosted by ShareMusic[1] and builds on the work of this institution as a Swedish national knowledge centre for artistic development and inclusion.[2]

## 1.2        A Short Excursion Into Modality

Multimodality is as old as human communication and probably older than human language. It is likely that human language emerged around 200,000 years ago, and that its appearance was accompanied by already evolved repertoires of facial expression, physical gesture and communicative "musical" intonation [1]. This means that all of the New London Group's "five modes" of communication - linguistic, visual, gestural, spatial and audio [2] - were most likely present among early human beings.

This process of evolution continued in all aspects of communication and art in all human cultures, including storytelling, theatre, song, opera and even spiritual life and worship. In general, these developments conformed to Gunther Kress's social semiotic approach [3] - as modes that could be shaped by culture, but retained intrinsic semiotic characteristics.

A change came in the late nineteenth century and accelerated throughout the twentieth century as technologies such as sound and film recording made multimodal experiences more readily and commonly available. And then came digital and immersive technologies, offering the possibility of intensive experiences of (in theory) many modes simultaneously, including opportunities for synaesthesia and for representations of objects and experiences in different modes, in such a way that one could substitute for another.

This had a particular impact on art and disability. First, there came a generation of music technology where spatial movement could replace touch, or touch could replace

---

spatial movement. Then, perhaps more importantly, came work on multimodal representations whereby those with visual impairment could "hear" paintings, or those who are deaf or hard-of-hearing could "touch" music, and deaf-blind people could both "see and hear" through haptic vibration and mapping.This is the essence of multimodality as understood in this collaboration.

### 1.3    Why a Knowledge Base for Arts and Inclusion?

Multimodal representations concern the description of information using two or more modes, combining text, sound, image, video and so one. Since we entered the age of digitization and even more so the Internet age, long-term preservation of digital objects describing real-world traces is on the agenda of many fields, particularly so for the information and computer sciences. Meanwhile we have standards (such as Open archival information system (OAIS) - ISO 14721:2012[3] [4]) and guidelines how to preserve data/documents in a FAIR[4] way [5] with the mean of trustworthy repositories [6].

Digital objects are complex. If it comes to traces of performing arts, examples are "'patch' files for processing what the performer plays" or "configuration files which map video capture of movement to musical performance" (examples taken from [7]). Digital repositories are equipped to deal with various types of digital objects, but how to best annotate them for human and machine consumption is at the forefront of a discourse about 'semantic artefacts' [8] and a synesthetic experience.

Indexing the content of a collection is a basic step in creating a repository or archive. It serves the information retrieval needs of potential users, but to some extent also represents selection criteria for building the collection in the first place. For the envisioned repository of artifacts around ShareMusic practices, describing the right objects in a right way, so that justice can be given to both the objects and their creators as well as to those who want to access them, has been central. One of the problems is that there is no easy, multipurpose, ready-on-the-shell classification to rely on. [9] This is why how to develop a **knowledge base for arts and inclusion** is at the core of this paper.

The term knowledge base is currently widely used. In its essence it refers to a way to create a shared system of insights. By using the term knowledge, the resonance is with humans being able to understand facts, to execute skills and to share beliefs and norms. In the computer sciences, the phrase knowledge base is connected to terms such as expert systems. Here the term knowledge is used to emphasize that there is more to a shared system than having a database, or an information management system. Although, if it comes to technical implementations these ways to digitally represent (ordered) knowledge are still around, be it databases or nowadays so-called knowledge graphs [10]. As we will describe later, when building a knowledge base techniques of ordering knowledge as studied by the field of Knowledge Organisation [11], and in particular its empirical, domain-analytics based approach become relevant [12].

---

[3] https://www.iso.org/standard/57284.html
[4] FAIR stands for Findable, Accessible, Interoperable and Re-usable

### 1.4    Overview of the Paper

This paper arises from the MuseIT project, which combines research with practical application. It presents a work-in-progress description, detailing how repository software can be developed to support accessibility and inclusion in the arts. The paper reflects the practical experiences with end-users and highlights the integration of scientific research and artistic practices.

The paper also explores how multimodality and accessibility are addressed, especially with the introduction of new AI technologies. These technologies drive the exploration of innovative solutions for better access to artistic content. It concludes with a summary of findings and discusses potential future research directions to enhance accessibility in the arts.

By investigating how multimodal needs can be met, this paper aims to contribute to the ongoing work of creating inclusive, accessible, and participatory artistic environments, allowing a broader range of people to engage with and enjoy the arts.

## 2      Interdisciplinary Collaboration, Co-creation and Long-term Archiving

As we sketched out in the introduction, art is a natural way to foster inclusion and has a long-standing history of forming alliances with new technologies. But scientific and societal progress is not only about ideas, it also needs actors (humans and institutions) and opportunities to bring them together. This section describes who has been involved in building this knowledge base for arts and inclusions. This way we aspire to bring the drive but also capacities of those involved in it to the foreground. Their paths in producing new knowledge and new experiences in the past informed what was chosen as a research question, and how the research question is approached.

If it comes to approaching inclusion and accessibility in a scientific manner various academic fields have a stake in this problem space, and interdisciplinary collaboration is inevitable. On the other side to enable the implementation of new ideas into the wider society needs institutions which act at the boundary between academia and society, and preferably have a record in co-creation and engagement with those which are 'target' of scientific research but at the same time the real owner of the challenges behind inclusion and accessibility. Such encounters across 'weak links' as long reaching connections are labelled in network theory require means. European project funding opens such areas, and the Horizon Europe call "Research and innovation on cultural heritage and Cultural and Creative Industries - 2021 (HORIZON-CL2-2021-HERITAGE-01)[5] created an opportunity for 6 projects in this area, among them MuseIT.

---

[5]    https://ec.europa.eu/info/funding-tenders/opportunities/portal/screen/opportunities/topic-details/horizon-cl2-2021-heritage-01-04https://web.archive.org/web/20250214113605/https://ec.europa.eu/info/funding-tenders/opportunities/portal/screen/opportunities/topic-details/horizon-cl2-2021-heritage-01-04

## 2.1    About MuseIT

MuseIT (Multi-sensory, User-centred, Shared cultural Experiences through Interactive Technologies) is a European Horizon project dedicated to enhancing accessibility and inclusion in cultural heritage experiences. Launched on October 1, 2022, and running until September 30, 2025.

"The primary objective of MuseIT is to co-design, develop, and co-evaluate multisensory, user-centered technologies that enriches engagement with cultural assets. This initiative focuses on individuals with disabilities, aiming to provide equal opportunities for all to access and participate in cultural and creative industries. By integrating advanced technologies and a participatory co-design approach, MuseIT seeks to democratize cultural experiences, ensuring that cultural heritage is accessible and engaging for everyone."[6]

The MuseIT consortium comprises nine partners from European Union member countries and three associated partners. This multidisciplinary consortium brings together organizations from cultural, technological, research, and user communities. It Includes partners DANS-KNAW, ShareMusic & Performing Arts and X-System Ltd, which are authoring this paper.

The project itself addresses three key challenges: *Accessibility of Cultural Assets*; *Broader Engagement with Cultural Assets and Cultural Co-creation*, *Methodologies for Preservation and Safeguarding of Cultural Heritage with Inclusion at Its Core*.

This paper focuses on the preconditions needed to enable better accessibility of cultural assets, in particular those emerging via co-creation and contributing to new methodologies to first document and second interact with cultural assets, takes the aspect of multimodality as its focal point. The work is carried out in the project's sixth Work Package, Multimodal Repository for Inclusion and Access. [7]

## 2.2    Introducing ShareMusic & Performing Arts

The Knowledge center ShareMusic & Performing Arts develops ways of working and methodologies that enable more people to experience, participate in and practice cultural and artistic activities. The forms of work include artistic labs, collaborative and research projects, lectures, performing arts productions, courses, workshops and other types of training and skills development, all centred around sharing knowledge of inclusive artistic practices. ShareMusic do not only share knowledge, but also constantly create new knowledge and are often part of interdisciplinary research projects and collaborations. The methods at use are co-creation and co-design processes where people with different backgrounds, experiences and abilities come together[13]. Materials that are created and collected range from scientific, to artistic and there between a lot of documentation of performances and best practice studies.

ShareMusic has always been at the forefront of technological developments, particularly in the field of music. An important part of the work is to explore the opportunities

---

for participation and creation that new technologies and digitization can offer to people with disabilities. Today much research is of course centred around how individuals with disability can be empowered through AI technology - and at the same time how these individuals' rights can be safeguarded. In this context, ShareMusic has a long-standing collaboration with X System, who developed the world's first functioning computational model of the musical brain [14]. This model can analyse musical tracks and predict with high accuracy the neurophysiological effects the music is likely to have on the listener, including autonomic activity, electrical brain activity, endocrine activity and neurotransmission. Many developments in X-System came through the experience work with young people with disability and children with trauma. X-System has wide experience in use of sensors and in both processing and representing data related to thoughts, feelings, the mind and the body.

To create a knowledge base was one of ShareMusic's motivations for joining MuseIT-project. We describe later along which dimension ShareMusic in this project pushes the boundaries, and how this relates to creating a knowledge base.

### 2.3     Introducing DANS-KNAW and the Dataverse project

DANS is an institute of the Royal Academy of Arts and Sciences (KNAW) and the Dutch Research Council (NWO). DANS roots go back to the former library of the Royal Netherlands Academy of Arts and Sciences and the Netherlands Institute for Scientific Information Services (NIWI-KNAW). Today, DANS has grown from an innovator in digital preservation of research data into the Dutch national centre of expertise and repository for research data. With a staff of now about 60 employees and about 300k datasets DANS belongs to the biggest repositories in Europe [15]. The main part of the collections DANS holds comes from the Social Sciences and Humanities. DANS has intimate knowledge in the field of cultural heritage [16]. Concerning the humanities, DANS provides a reference archive service for the archaeology community commissioned and guided by the Cultural Heritage Agency. DANS is engaged in many European projects, providing expertise in Research Data Management, FAIR data, and exploring new technological solutions. To emphasize the strong community connection of the DANS services [17]. DANS has organised its services into so-called Data Stations of which one is dedicated to the Humanities. Content-wise, most of the collection items are text-based, but images and visual-audio material is also part of the collection. Most notable projects DANS has been involved and which also informed DANS contributions to MuseIT is the SSHOC project[8] in which a user-friendly way to deploy archival (web-based) services was developed [18]; the Polifonia project[9]; and the ongoing projects FAIR-Impact[10] and SSHOC-NL. In the latter two projects new ways to annotate digital objects [21] and interfaces (the ODISSEI (Open Data Infrastructure for Social Science and Economic Innovations) Portal [19,20]) were developed which informed this paper.

---

[8] Social Sciences & Humanities Open Cloud - SSHOC. EU funded Project ID 823782. 2019-2022
[9] Polifonia: a digital harmoniser for musical heritage knowledge. EU funded, Project ID 101004746.
[10] FAIRImpact project. EU funded, Project ID 101057344.

DANS started with web-based services early on and nowadays operates various Data Station services based on the Dataverse Platform. The Dataverse project[11] is an open source, community-oriented project, and has been initiated by Harvard IQSS - an institute with comparable collections to DANS origins. Since its beginning Dataverse has grown into a universal- all fields repository software application, which is rather agile and adaptable to the needs of various clients, while providing regular updates of the core technology [22]. In its original design Dataverse carried a strong emphasis on the users being able to shape their own 'data sharing environment' almost like a Virtual Research Environment. At the same time, Dataverse instances can be run in a way that enables centralised authoritative data curation by the managers of the archive. This volatility makes it an excellent platform to experiment with new technical features growing out of both new user needs and new technological opportunities. We will discuss later, how this ability has been used for the topic of this paper.

## 3    A Knowledge Base for Arts and Inclusion – Conceptual Challenges and Possible Interactions with Users

### 3.1    The Need for a Unique Resource

As mentioned, there is no single, comprehensive knowledge base implemented into a repository, anywhere in the world, both serving the needs of disability and the creative and performing arts and offering access to related cultural assets. These needs include scientific and humanistic research and development of artistic practice and technology. ShareMusic, in its role as knowledge centre, has for long aspired to create such a resource and the MuseIT project, through the development of multimodal approaches, presented the opportunity to develop it together in the relevant partnership with DANS and X-System.

At the same time the aspiration is to create a repository - a digital archive for artefacts which represent research leading to new ways of inclusion and accessibility, but also to develop a multimodal archival process for new products/outcomes of new inclusive practices in the arts, specifically within music and the performing arts. The envisioned users of both knowledge base and archive will be people with disability, artists, organisations serving arts and disability, organisations serving culture and the arts, organisations serving disability, independent activists, special needs centres and organisations, hospitals, nurses and carers, support workers and animateurs, individual researchers, university and college arts departments, rehabilitation departments and departments concerned with disability and inclusion. Hence the aspiration has from the beginning been cross-sectoral and this also creates specific challenges for the knowledge base. The knowledge base will offer, in terms of solution, product, service and process, the following:

---

[11] https://dataverse.org/

1. Access to abstracts/publications, humanistic and scientific, related to arts, disability and inclusion. This will be achieved in the digital domain, including appropriate ontologies and knowledge graphs.
2. Access to unpublished materials which will be aggregated manually, digitised and ported to the repository.
3. Examples of good practice, both mono- and multimodal, aggregated both digitally and manually, where necessary digitised, and then ported to the repository.
4. Manuals, toolkits and descriptions of process aggregated both digitally and manually, where necessary digitised, then ported to the repository.
5. Multimodal experiences in cultural assets made available through digital and manual aggregation, through other parts of the MuseIT project such as the Remote Performance Platform and through the MuseIT repository standard.

The core of the newly to be built archive are resources ShareMusic and its collaborative network already had documented in digital form, but not put together at one platform, nor annotated based on a knowledge base. Much also comes from grass root organisations, within the performing arts sector, that have been "carrying the banner" of art and inclusion but with no organisational tradition of documenting and archiving its practises or cultural outputs.

As part of MuseIT, ShareMusic also leads a Work Package which is concerned with the development of a Remote Performance Platform, utilising low latency technologies to facilitate co-creation services for creating born digital content. The content produced from this Work Package is supposed to form part of the collection. Music is an essential part of cultural heritage and one of the most challenging to transmit multimodally. Music can consist of tangible cultural heritage such as notation, instruments, and recordings. To document the traces of musical heritage based on shared, shareable and machine readable knowledge graph technology has been proven very challenging in itself [23,24]. However, a music performance is an intangible heritage. This is why music is an integral part of MuseIT. If we are able to create multimodal experiences of music, we will be able to do it with other art forms as well. The user-centred and co-creative approach of the MuseIT project allows for exploring how it feels to engage in creating art as well as experience the art.

### 3.2    Utilising a data repository as a Knowledge Base

According to WHO, globally, more than 2.5 billion people need one or more assistive products. In many countries, most people who need assistive technology do not have access to it.[12] Related to increasing accessibility of the physical world is the work of increasing accessibility of the online world, largely guided by the resources and guidelines developed by the World Wide Web Consortium's Web Accessibility Initiative (W3C WAI[13]).

Moving into the web accessibility space, the W3C WAI guidelines have been in production or discussion since 1994, with the first set of guidelines finalised in 1999, specifically related to Web Content Accessibility Guidelines (WCAG) that relate to web-based content and the user experience. Other guidelines include the User Agent Accessibility Guidelines (UAAG), which refers to web browsers and media players (as well as assistive technologies), and the Authoring Tool Accessibility Guidelines (ATAG) that inform the infrastructures of web development and content creation. These three guidelines are non-normative and focus on how user experiences are experienced, while more technically focussed guidelines exist for specific tools and programming tools. The ongoing development of the WAI guidelines reflect the changing nature of web design and the internet, as well as new technologies. The fundamentals of European legislation like the European Web Accessibility Directive refer to the standard EN 301 549, which in turn clearly references the WCAG 2.1 guidelines along with extra requirements. Harmonization at the legislative level means that the WCAG guidelines are updated faster than the legislation is; the WCAG 2.2 guidelines have since been approved and recommended beyond European legislation and the WCAG 3.0 are currently under community discussion.

---

[12] (https://www.who.int/news-room/fact-sheets/detail/assistive-technology)
[13] (https://www.w3.org/WAI/)

Dataverse with its specific open software collaboration nature has developed many functions and is continuing to do so in many directions simultaneously. As described above, one of those functions is to provide a data management system for researchers and academic communities to share data openly. The interface of Dataverse is built to cater for such communities and a certain pre-knowledge of how to use repositories, in particular how to provide the metadata content when depositing data is needed. Since the **Knowledge Base for Arts and Inclusion** is meant to serve a greater audience than just research communities, one part of the project development so far has been concerned with building an accessible and intuitive front end on the ShareMusic Dataverse instance.

The front end of the Knowledge Base for Arts and inclusion currently consists of a few accessible frames, where the user can find, interpret, and use the materials in different multimodal modes (see Fig 1). All data and metadata will be stored in Dataverse and retrieved through API calls. We are also working with other groups who have been involved in previous projects where a Dataverse application have been used similar capacities (e.g. the Now.Museum). Since the front-end application is what most end users will meet, a reference group is being formed to help us evaluate the functions and accessibility features of the front end. The reference group will consist of disabled artists, with physical, sensory and intellectual disabilities and of cultural workers and teachers wanting to learn more about art and inclusion. In the final months of this still ongoing project, there is time dedicated for an iteration process where designers and developers can alter things dependent on the reference groups input.

In order to ensure web accessibility standards are met, the co-designed front end of Dataverse will also be tested with automated and manual tools, add-ons, extensions and bookmarklets.

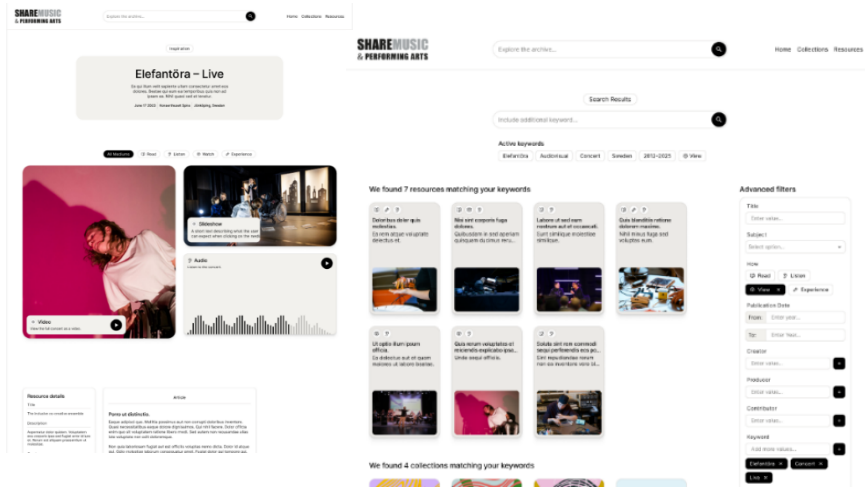

**Fig. 1.** A snapshot of the mock-up wireframes the design of a record and a search page in the Knowledge Base for Art and Inclusion can look.

The lack of web accessibility training in computer studies and information studies has been documented for more than a decade, with all reaching the conclusion that while comprehensive web accessibility training is necessary for developers to understand and implement the guidelines discussed (as well as understanding why they are important), there is very little focus on web accessibility within the curricula of these programs [25, 26,27]. Nonetheless, we will in this project also train artists and staff, from the performing arts and inclusion field, with no prior academic background in the repository management of the specific Dataverse Instance. For that reason, we have developed a Registration Guideline for the Repository Management of Multimodal Data. The idea is also to make Dataverse more accessible by simplifying the usage, listen to the feedback we receive and highlight that everyone can archive. The hope is that this will make individual artists with disabilities as well as organisations better in preserving and disseminating their cultural outputs, their knowledge and their good examples of best practice. By this we aim to extend our community and create an iterative process where both the repository (Dataverse) and the Knowledge Base (the front end) are continuously developed based on user needs.

### 3.3   Accessibility, Multimodality and AI

So far, we have explored the role of multimodal accessibility and the challenges of documenting and archiving multimodal content in an inclusive and structured way. As digital repositories evolve, AI plays an increasingly important role in bridging accessibility gaps and ensuring content remains adaptive and meaningful for diverse users.

To address these challenges, we introduce two distinct AI-driven approaches, Monomodal Transformative AI (MTA), a group of technologies that convert a single-source input into multiple accessible formats, and Multimodal Cognitive AI (MCAI), a group of technologies trained on multiple modalities to generate context-aware outputs utilizing multimodal knowledge bases. While MTA is already widely applied in accessibility solutions, MCAI remains in its early stages and will emerge alongside the development of multimodal data stores. Both approaches hold potential for deeper contextual adaptation and will shape how multimodal repositories evolve, improving data structuring and content accessibility.

**Expanding Accessibility Through AI**. Integrating multimodal AI into a Knowledge Base for the Arts and Inclusion can enhance accessibility by transforming original content into diverse formats. MTA encompasses technologies such as text-to-speech, braille conversion, captions, and haptic feedback, ensuring that users can engage with content in a format suited to their needs.

MCAI, though still in early stages, extends accessibility by reasoning across diverse inputs from multiple sources. Rather than relying solely on text-based transformations, MCAI can infer context from various modalities, enabling a more natural and fluid flow of information. Over time, MCAI holds the potential to enhance the interpretation of cultural assets, making data more accessible and meaningful from a user-centred perspective.

For an ideal implementation, multimodal repositories should integrate both MTA and MCAI methodologies. While MTA is already well-established and impactful, MCAI will continue to evolve alongside advancements in multimodal data storage and interaction design. Future large language models (LLMs) may leverage these repositories for more advanced learning and contextual adaptation, but at present, MTA remains the foundational tool for structuring and enriching databases to support accessibility.

Presently, Monomodal Transformative AI (MTA) technologies serve as the primary means of ensuring accessibility, making it crucial to manage their limitations effectively [28]. Errors in text recognition can distort speech synthesis, captions, or braille output, and without verification, these inaccuracies may propagate. Additionally, MTA must be applied with care and context-awareness, ensuring its transformations are appropriate and do not introduce distortions. Not all AI-driven transformations enhance accessibility, excessive processing or abstraction can strip key details, reducing clarity rather than improving usability. Effective filtering, validation, and tracking of multimodal transformations are essential to maintaining data integrity and ensuring meaningful access to cultural assets.

As Multimodal Cognitive AI (MCAI) technologies emerge, they will remain highly dependent on MTA outputs. If flawed data is used for training, MCAI may reinforce inaccuracies, leading to unreliable reasoning and accessibility gaps. While future language models (LLMs) may adapt contextually, MTA remains the essential tool for structuring multimodal databases and ensuring accessibility today [29]. Accurate and contextually appropriate transformations are critical to supporting the long-term usability of multimodal repositories, particularly in facilitating the documentation and interaction with cultural assets in inclusive environments.

### 3.4   Case Studies of Interaction with (mono)Material for Increased Digital Accessibility

MTA is the most researched and widely implemented approach in multimodal communication. Text-to-speech, speech-to-text, and alternative text descriptions for images are well-established accessibility tools, supported by extensive research and integrated into both European and national legal frameworks.

Within the cultural heritage domain, including museums, theatres, and community creation spaces, recent studies into the presence of accessibility for disabled artists and users support co-creation initiatives for increasing accessibility but are often limited to the physical aspects of the space in question.

A study detailing two Portuguese museums and the intentional multimodal design of their exhibits looked at an "access for all" approach that included discourse and improvement with potential patrons with lived experience of disability [30]. The two museums in question worked alongside museum curators, technicians, and disabled patrons on approaches that added layers of modality to the exhibits within the physical space of the museum (including 3D printed models, touch-based exhibits, audio guides, and Braille item descriptions). However, this approach is firmly rooted in the physical - the museum exists in a physical space and further (digital) representations of the collections and how they could be experienced were not part of the study, likely due to the technological limits at the time.

More recently, a study into a community music initiative (Music Community Lab) that intentionally works in the music technology realm: hosting music hackathons that aim to solve problems put forward by community members across various modalities (visual, haptic, audio, etc.) [31]. The focus of the study was to determine how the hackathon could be made more accessible to community members, but still in a way that focussed on access to physical space or accommodations provided within the venue, not necessarily related to the accessibility of the technology used in the hackathon, which is to be expected given that the question of the study was related to the events put on by the MCL being accessible, rather than the methods of technology within the MCL hackathons being accessible to enable co-creation.

Therefore, the focus from the performing arts as well as the cultural heritage at the current time is generally on the physical space and the events or engagement that the community is a part of, rather than the technology that it uses or the web-based accessibility of the technologies. Thereto, the outputs created within the Performing Arts sector can carry other meanings and artistic expressions than those of museum objects. As such, the multimodal interpretations benefit from other type of interactions than those linked to musealised cultural assets. Performing Arts, such as theatre, dance and opera are multimodal meaning-making events in themselves. They include many expressions such as language, music and body movements and are often intangible [32]. When the Performing Art is being archived the act of interpretation and documentation is always very present. Records of performances have many purposes, including being factual recordings or attempts to capture and inspire an audience.

By using Dataverse as the repository, an open model encouraging multiple representations allows for creative reuse and reinterpretation of the assets. It keeps the spirit of the performances alive and is envisaged as the future of the performing arts archive - opening it up for reuse [33]. Dataverse creates opportunities in the Knowledge Base front end for multiple presentations of one resource.

To ensure convenient access to all resources for all user types, especially individuals with disabilities, the team has developed the LLM-powered "Ask ShareMusic" service[14]. This service enables users to create separate collections for various multimodal material types, such as news, publications, papers, audio, video, and data preserved in the ShareMusic Dataverse repository. Users can interact with the service by asking questions through both voice and typing. MuseIT has integrated all resources using the Croissant for Machine Learning[15] [21] standard, developed by a consortium of industry and academic partners, including DANS and Kings College London.

### 3.5 Developing Structured Repositories: Development in Dataverse to Register Multimodally Born Materials

It is possible to argue that many archives and museums were originally built for the tangible and musealized artefacts and data. Since Intangible Cultural Heritage was recognised in the last decades and UNESCO developed instruments to safeguard and manifest its status, attempts have been made to develop and alter current archival management systems to fit more kinds of documentation materials[16].

The documentation, reproduction and presentation of performing arts and other multimodal intangible cultural expressions in general, benefits new technological methods and models for data curation. Therefore, we have developed a multimodal metadata schema for Dataverse to support different types of materials deposited by users and collected and registered outside of our repository [34]. The rapid development of Machine Learning tools and frameworks allowed us to build innovative pipelines and apply multimodal transformations which are being deposited and archived back in Dataverse, with keeping provenance chains and providing metadata enrichment with external controlled vocabularies.

Important factor of successful development was also work on the Croissant for Machine Learning standard(link) where DANS collaborated with other MuseIT partners on the specification for the Machine Learning tools integration with metadata records. While enabling this new multimodalities metadata schema, users are getting new fields in the metadata block in Dataverse representing modalities (see Fig 2).

---

[14] https://ask.sharemusic.se
[15] https://docs.mlcommons.org/croissant/docs/croissant-spec.html
[16] https://www.unesco.org/archives/multimedia/?pg=14

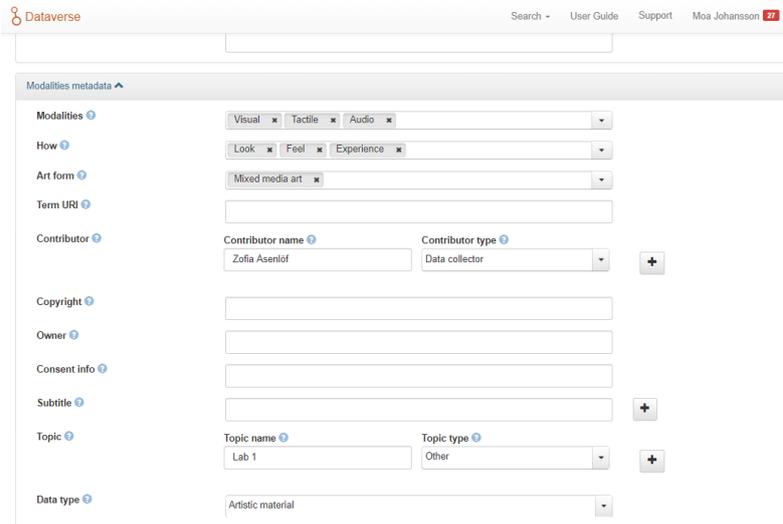

**Fig. 2.** Snapshot of the Modalities metadata block (a specific extension to the Dataverse core citation block) from the ShareMusic Dataverse instance.

After the metadata record is saved, it is exported in various formats and available both for search and harvest by search engines and being ingested in the ShareMusic knowledge graph accepting JSON-LD serializations. This allows more advanced metadata integration with external tools where users can switch multimodal representations in user interface based on their preference. Additionally, since metadata is available in the knowledge graph on a very detailed level, it allows linking the record with other collections providing extensive view from the artistic point of view, for example, to show all performances of some artist.

Dataverse as data archiving technology also plays a role of "bridge" to Artificial Intelligence applications recognising the type of modality and connecting AI service allowing processing, extraction and adding extra data levels, for example, to do speech to text recognition for audio files and image to text recognition. This approach allows us to connect different types of AI applications in the pipeline called Agentic, with the possibility to use output of one AI app as input for another AI app in the processing chain. To illustrate the approach, we generate screenshots in the video registered in Dataverse and using it as input for the computer vision service detecting all objects and people on the screenshot and using it as input for the text to speech service describing the image (see Fig 3).

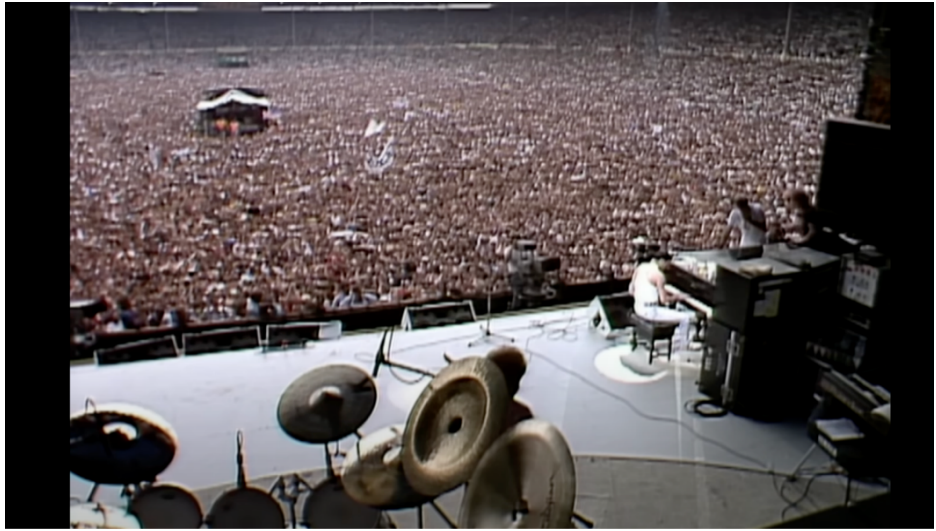

**Fig. 3.** Queen performance during Live Aid in 1985[17]. Description delivered by AI-powered computer vision service: "*there is a large outdoor concert with a massive crowd in attendance. The stage is visible in the foreground, featuring musical instruments such as a drum set and a piano. There are two musicians on stage, one playing the drums and the other playing the piano. The crowd extends far into the distance, filling the entire viewable area. The atmosphere appears lively and energetic, typical of a large music festival or concert.*" (Image from Youtube)

## 4     Beyond Static Archives: Multimodal Registration, Reenactment and AI-Driven Accessibility

### 4.1     Multimodal Registration for Accessibility and Community Engagement

Building on insights from the MuseIT project, this section explores new frontiers in multimodal accessibility, knowledge representation, and audience engagement, advancing from static archives to dynamic, participatory ecosystems. By leveraging emerging AI methodologies and co-creation frameworks, this research reimagines how accessibility, artistic interpretation, and public engagement can evolve in multimodal spaces.

As has been previously explained, traditional archiving is often static and text based, but performing arts and multimodal registration allows accessibility to evolve through co-creation and participatory engagement. By integrating community-driven methodologies, multimodal content is not only preserved but continuously enriched by diverse contributors. These reinterpretations become part of the evolving archive, ensuring that artistic expressions remain accessible, reflective of contemporary culture.

---

[17] https://www.youtube.com/watch?v=vbvyNnw8Qjg

Unlike conventional documentation, which primarily records artifacts and documentation of performances, multimodal registration could capture the interaction between users, artists, and the digital space. This creates knowledge ecosystems where experiences can be rediscovered, reinterpreted, and layered with new meaning over time. As art forms evolve, this process enables new creative expressions to acknowledge the past while shaping the future, ensuring an ongoing dialogue between past and present artistic practices.

## 4.2    The Remote Performance Platform as an Attempt to Explore Modalities Beyond Conventional

In the MuseIT project, we explore this unconventional documentation through the development of a Remote Performance Platform. The platform will enable music co-creation at a distance. Not only will musicians be able to explore the sound of music but explore the feeling of music.[18] By including the communication of thoughts and emotions between participants, we are developing a remote co-creative artistic process which we will try to document in the Knowledge Base for Art and Inclusion.

Advancements in cognitive science and multimodal AI introduce unconventional pathways for accessibility, interaction, and interpretation. Physiological data, such as EEG, heart rate variability, and skin conductance, can infer emotional states, allowing AI-driven systems to dynamically adjust how multimodal content is displayed or experienced.

Similarly, AI techniques can extract contextual meaning from images, facial expressions, and body posture, enhancing emotional mapping across multiple modalities. These insights could inform adaptive systems, modifying environmental factors, such as lighting, sound, or tactile feedback, in response to user needs. By dynamically adapting to emotional and cognitive states, these systems create personalized and responsive accessibility experiences, ensuring a more intuitive and immersive interaction with multimodal content.

Looking forward, archiving audience responses over time can reveal evolving cultural perceptions, providing a historical record of how artistic experiences have been interpreted across different contexts. This not only enhances accessibility but also offers a unique insight into the evolution of artistic engagement, ensuring a richer, data-informed approach to multimodal preservation.

## 4.3    Re-enactment of Living Archives for Public Audiences

Multimodal archives capture more than static records—they preserve the lived experiences and emotional dimensions of cultural heritage. When emotional states are recorded as part of multimodal datasets, they allow deeper contextualization from both the performer's and audience's perspectives. It gives insight to the co-creative process as such.

---

[18] https://www.sharemusic.se/resources-and-inspiration/museit

We propose developing new strategies for the 're-enactment' of living archives, enabling audiences to engage with historical performances, cultural expressions, and co-created artifacts in immersive and dynamic ways. Potential approaches include AI-driven reconstructions, multimodal storytelling, and adaptive content generation that responds to user interactions in real time, transforming archives from passive collections into interactive, evolving spaces.

By rethinking how multimodal data is processed and experienced, this research could to expand the creative and expressive potential of accessibility technologies while ensuring ethical considerations and data permissions are upheld. And as with all developments carried out jointly by the authors of this paper, with a co-creative and co-design methodology guiding the research.

## 5 Findings and Prospects for Further Research

This paper discusses the creation of a Knowledge Base for Arts and Inclusion andr address the lack of a comprehensive platform integrating arts, disability, and inclusion. It examines how new technological tools can strengthen digital inclusion, focusing on the role of multimodality in arts accessibility and participation. The research centers on a repository for archiving performing arts artifacts and improving accessibility for people with disabilities.

The Knowledge Base aims to preserve and disseminate cultural outputs from disabled artists and organizations. By enhancing the Dataverse systems functionalities, it empowers users to share their works, best practices, and knowledge. The project integrates AI technologies—Monomodal Transformative AI (MTA) and Multimodal Cognitive AI (MCAI)—to increase accessibility by transforming content into diverse formats. However, the paper highlights the need for careful management of MTA technologies to ensure context-aware transformations.

The project also introduces multimodal registration, where content is continuously enriched through community-driven engagement, creating a dynamic knowledge ecosystem. This approach transforms the archive into a participatory space that evolves over time. Through technologies and collaboration with disabled artists and cultural workers, the project aims to enhance the preservation, sharing, and accessibility of artistic expression.

The paper concludes, with a look at three prospects for further research, that:

— development and refinement of MCAI could significantly enhance the platform's ability to provide context-aware outputs across multiple formats;
— there is an opportunity to explore how multimodal archives impact user experience and engagement. Research could investigate how these types of archives influence learning, memory, and emotional engagement in users with disabilities, as well as how they can support more inclusive cultural practices;
— continued participatory research with disabled artists, cultural workers, and other stakeholders is essential to ensuring the platform meets the needs of diverse users. Involving a wide range of individuals with different disabilities will enable iterative design improvements. This may include studies on the most effective ways for users to interact with multimodal content, the effectiveness of various assistive technologies, and how to improve ease of use for individuals without academic backgrounds in repository management.

**Acknowledgments.** The research and the developments of a Knowledge Base for Art and Inclusion have been made possible by the MuseIT project, coordinated by Nasrine Olson at Högskolan i Borås. MuseIT is Co-funded by the European Union (Project ID: 101061441). The work presented in this paper also builds on projects such as Polifonia: a digital harmoniser for musical heritage knowledge (EU funded, Project ID 101004746), and SSHOC.NL - a Dutch funded research infrastructure project in the Social Sciences and Humanities. We would like to thank partners from the Now.Museum initiative (https://www.now.museum/now-museum/) for their ongoing support. Part of the paper has been informed by participation in the FAIRImpact project (EU funded, Project ID 101057344)

**Disclosure of Interests.** The authors have no competing interests to declare that are relevant to the content of this article.